\documentclass[onecolumn,3p,preprint,a4paper,10pt]{elsarticle}
\usepackage{amssymb,natbib,amsmath,wasysym}
\usepackage{multicol}
\usepackage{caption}
\usepackage{color}

\renewcommand{\d}{\mathrm{d}}

\newcommand{\kpc}{\mathrm{kpc}}
\newcommand{\Gyr}{\mathrm{Gyr}}

\newcommand{\muG}{\mathrm{\mu G}}

\newcommand{\divb}{ {\bf \nabla}\cdot{\bf B}}
\newcommand{\divA}{ {\bf \nabla}\cdot{\bf A}}

\newcommand{\divv}{ {\bf \nabla}\cdot{\bf v}}
\def\divB{\divb}

\def\bfA{\mathbf{A}}
\def\bfv{\mathbf{v}}
\def\bfB{\mathbf{B}}

\def\curl{ {\mathbf \nabla}\times}

\def\be{\begin{equation}}
\def\ee#1{\label{eq:#1}\end{equation}}
\def\bea{\begin{eqnarray}}
\def\eea#1{\label{eq:#1}\end{eqnarray}}
\def\bee{\begin{equation}}
\def\eee#1{\label{eq:#1}\end{equation}}

\def\EQ#1{Eq. (\ref{eq:#1})}

\newcommand\bfn{{\mathbf \nabla}}
\newcommand\bfV{{\mathbf V}}

\newcommand{\FIG}[1]{Fig. (\ref{fig:#1})}
\newcommand{\FIGp}[1]{Fig. \ref{fig:#1}}
\newcommand{\bFIG}{\begin{figure*}}
\newcommand{\eFIG}[1]{\label{fig:#1}\end{figure*}}
\newcommand{\bFIGs}{\begin{figure}}
\newcommand{\eFIGs}[1]{\label{fig:#1}\end{figure}}

\begin{document}

\title{A Vector Potential implementation for Smoothed Particle Magnetohydrodynamics}

\author{Federico A. Stasyszyn}
\ead{fstasys@aip.de}

\author{Detlef Elstner}
\ead{delstner@aip.de}

\address{Leibniz-Institut f\"ur Astrophysik Potsdam (AIP), An der Sternwarte 16, 14482, Postdam, Germany}

\begin{abstract}

The development of smooth particle magnetohydrodynamic (SPMHD) has significantly improved 
the simulation of complex astrophysical processes. 
However, the preservation the solenoidality of the magnetic field is still a severe problem for the MHD. 
A formulation of the induction equation with a vector potential would solve the problem. 
Unfortunately all previous attempts suffered from instabilities.  
In the present work, we evolve the vector potential in the Coulomb gauge and smooth the derived magnetic field for 
usage in the momentum equation. 
With this implementation we could reproduce classical test cases in a stable way.  
A simple test case demonstrates the possible failure of widely used direct integration of the magnetic field, 
even with the usage of a divergence cleaning method. 

\end{abstract}

\maketitle


\section{Introduction} \label{sec:intro}

In recent years, not only the presence but the morphology of magnetic fields in galaxies has been determined \citep{1996ARA&A..34..155B}.
It represents a huge scientific challenge, as this is a new opportunity to understand 
how the magnetic field is related to the astrophysical hosts, their history and properties.

A possible explanation for the magnetic field amplification is the action of a dynamo driven by turbulence 
and large scale gas motions \citep{2002RvMP...74..775W,2009ASTRA...5...43B,1988ASSL..133.....R}, where 
the unknown initial seed field is washed out by the turbulent character 
of the flow. Numerical simulations of evolving galaxies should help to understand the main properties of the magnetic field 
amplification with the observed morphology. 

The success of cosmological simulations using SPH methods motivates the application of that technique 
also for the MHD case \citep{DolagStasyszyn2009}. 
The direct implementation of the induction equation with the magnetic field unfortunately 
suffers from the preservation of solenoidality.
The artificial growth of $\divB$ in these schemes is usually reduced by a more or less artificial cleaning of $\bfB$.
The direct integration of $\bfB$ with or without cleaning may lead to unrealistic numerical growth of 
the magnetic field as it occurs in the example described in section \ref{sec:test} or in \citet{Kotarba2009}.   
There is no $\divB=0$ preserving scheme known for SPMHD integrating the magnetic field $\bfB$ directly from the induction equation. 
Changing the integration variable from the magnetic field to the vector potential $\bfA$ with $\bfB = \curl \bfA$ solves the problem in a natural way. 
A vector potential formulation in SPH was previous studied in detail by \citet{PriceVector2010}. 
The implementation was working for one and two dimensional problems, but failed
in three dimensions.

In the following sections we present an application of the vector potential in SPMHD, which overcomes the previous problems. 
In section \ref{sec:impl} we shortly describe our implementation, followed by the analysis of some test cases in section \ref{sec:test}. 
Finally we discuss possible implications 
in section \ref{sec:disc} and we present our conclusions in section \ref{sec:conc}.

\section{SPH Implementation}\label{sec:impl}

Throughout this work we will use the SPMHD version of {\small Gadget-3} \citep{StasyszynDolag2013},
where the ideal MHD is solved following the induction equation in the form
\begin{eqnarray}
	\frac{\d \bfB}{\d t} &=& (\bfB \cdot {\bf \nabla}) \bfv - \bfB ({\bf \nabla} \cdot \bfv)
\label{eq:induc0} 
\end{eqnarray}
in which, we assume the $\divb=0$ constraint is valid, by
taking special care on reducing it \citep{StasyszynDolag2013,Tricco2012}.

\citet{PriceVector2010} studied carefully the possible SPH vector potential formulation,
we follow it and use it as a starting point. 
The definition of the magnetic field and the evolution of the vector potential can be summarized as follows:
\bea
\bfB &=& \curl \bfA \\
\frac{\d \bfA}{\d t} &=& \bfv \times \curl \bfA + (\bfv \cdot \bfn) \bfA - \nabla \phi 
\eea{Ainduc0}
where $\phi$ is an arbitrary scalar representing the freedom to choose a special gauge.
There is the freedom of choosing different gauges for each time step if desired to improve the numerics, 
but keeping track of a proper $\phi$ evolution of a given particular gauge. 

In tensor form the components of \EQ{Ainduc0} simplify to
\begin{eqnarray}
	\frac{\d A_i}{\d t} &=& v^j \frac{\partial A_j}{\partial x^i} - \frac{\partial \phi}{\partial x^i}
	\label{eq:AEvol0}
\end{eqnarray}
where $i,j$ are component indexes and summation over double indices is used.

In the SPH framework this equation is written as follows,
\begin{eqnarray}
\frac{\mathrm{ d}A_a^i}{\mathrm{ d}t} &=&
\frac{ f_a}{\rho_a} \left[\sum_{b=1}^{N} - m_b 
\left(\phi^i_{ab} - 
v_a^jA^j_{ab} \right)\partial W^i_{ab} \right]
\end{eqnarray}
where $a,b$ are particle indexes, $f_a$ is the correction factor that arises from the use of variable smoothing lengths, 
the $A^j_{ab}$ is the difference between the potential of neighboring particles and $\partial W^i_{ab}$ is 
the kernel gradient operator between particles (for more details refer to \citet{DolagStasyszyn2009}).

As we mentioned before, the gauge choice does not manifest in the magnetic field, but in the evolution of the vector potential.
For example, if we use the {\it Coulomb} gauge, which means $\divA = 0$ for all points of space and time, 
we have to take care of fulfilling this requirement. 
Therefore, we face a similar problem as keeping $\divB = 0$, that has already been extensively studied 
\citep{StasyszynDolag2013,Tricco2012}. 
We take a similar approach, using a cleaning scheme \citep{Dedner02} originally thought to lower the $\divB$ errors, 
but applied to $\bfA$ in order to ensure $\divA=0$. 
The solution of the problem is equivalent to choose a modified {\it pseudo-Lorenz} or {\it velocity} gauge \citep{2002AmJPh..70..917J,2007NJPh....9..305B}, 
 with an additional damping term.
Note, that keeping $\divA = 0$ will also simplify the calculation of the diffusion 
terms for the non-ideal MHD equations. 
The evolution of the gauge is achieved through following equations
\bea
\frac{\d \phi}{\d t} &=&  - c_h^2~\divA - c_h\frac{\phi}{h} - \frac{\divv~\phi}{2} 
\eea{ADCS0}
where $c_h$ is the characteristic signal velocity, $h$ is the smoothing length and we add the final term, 
introduced by \citet{Tricco2012} that takes into account compression or expansion of the fluid.
\citet{Tricco2012} found that this additional term, improves conservation of energy and in particular for 
the divergence cleaning is crucial the symmetrization of the SPH operators. 
In our case we use a ``differential'' non-symmetric SPH operator and we do not apply any {\it limiter} 
as in \citet{StasyszynDolag2013}, and seems sufficient to achieve stability. 
However, when coupling the energy evolution using a symmetric operator can improve the energy conservation.

Therefore the gauge evolution in SPH form writes as \EQ{ADCS0}, 
and the differential operators takes the form for the divergence case as:
\bea
\divA &=& \frac{ f_a}{\rho_a} \left[\sum_{b=1}^{N} m_b A^i_{ab}\partial W^i_{ab} \right]
\eea{DivA}

\subsection{Lorentz Force}\label{subsec:force}

The coupling of the magnetic field and the momentum equation is done by calculating $\bfB$ from the potential and 
then applying it to the force equation using the magnetic stress tensor \citep{1985MNRAS.216..883P} 
\begin{equation}
M^{ij} = \left(\frac{B^i B^j}{\mu_0} - \frac{1}{2\mu_0}|\bfB|^2\delta^{ij}\right) .
\label{eq:maxtens}
\end{equation}
which modifies the momentum equation by adding the term 
\begin{equation}
\left(\frac{\d v^i}{\d t}\right)_{mag} = \frac{1}{\rho} \left( \frac{\partial M^{ij}}{\partial x^j} - B^i \frac{\partial B^j}{\partial x^j} \right)
\label{eq:momen}
\end{equation}
The last term subtracts  $\divB$ errors which still occur here by the numerical approximation of the 
curl operator. It stabilizes the numerical scheme \citep{2001ApJ...561...82B}.
The use of this set of equations corresponds to the hybrid approach described by \citet{PriceVector2010} and 
is equivalent at the force treatment described in \citet{DolagStasyszyn2009} and subsequent works.

In order to further stabilize the scheme  $\bfB$  we smooth magnetic field $\bfB$ before it is applied in the 
calculation of the Lorentz force. 
In contrast to the implementation presented by \citet{DolagStasyszyn2009}, 
we do not volume weight the smoothing operation, because it generates instabilities in low density regions. 
We smooth the field of the neighboring particles using only the SPH kernel computing
\begin{equation}
B^i_a = \frac{\sum_b B^i_b W_{ab}}{\sum_b W_{ab}}.
\label{eq:smooth}
\end{equation}
Note that this additional step does not introduce any artificial dissipation for the induction process, 
because we do not change the time evolution of the vector potential.
We use the same switches and the magnetic signal velocity described in \citet{DolagStasyszyn2009}. 
To summarize, we calculate a $\bfB$ from the potential $\bfA$, afterward we smooth the magnetic field 
and calculate the corresponding forces, using the same numerical corrections from previous implementations. 

\subsection{Diffusion}\label{subsec:diff}

It is straightforward to implement a diffusion of the magnetic field through the potential. 
Taking care of keeping $\divA=0$, we may simply write the diffusion term as
\begin{equation}
	\left(\frac{\d \bfA}{\d t}\right)_{diff} = \eta_i \nabla^2 \bfA
	\label{eq:ADiff}
\end{equation}
Note that, this formulation allows a spatial dependent dissipation $\eta$. 
The Laplacian calculation as described  
has previously been successfully implemented for spatially independent diffusion of the magnetic field \citep{Bonafede11,Price2010}, 
and has been proven to be useful handling possible numerical instabilities, and is written in SPH\footnote{Eq. 3 in \citet{Bonafede11}  as 
\bea
\left(\frac{\d A_i}{\d t}\right)_{diff}
 &=& 2 \eta_i \rho_a \left[\sum_{b=1}^{N} \frac{m_b}{{\hat{\rho_{ij}}^2}} A^i_{ab} \frac{\mathrm r_{ij}}{|\mathrm{r_{ij}}|}\partial W^i_{ab} \right]
\eea{DifSPH}
}

\section{Testing the method}\label{sec:test}

We first show a simple kinematic test, where the direct integration of the magnetic field fails 
in contrast to the vector potential formulation of the induction equation.
In order to apply this formalism for astrophysical problems, the non-linear effects due to feedback 
from the magnetic field on the dynamics have to be successfully treated in different environments.  
Therefore, we present the classic shock tube from \citet{BrioWu88} 
and  the more complex example of the \citet{Orzang79} vortex with the vector field implementation. 
We always perform the tests in 3 dimensions, because previous attempts \citep{PriceVector2010}
failed particularly in those cases. 

\subsection{Differentially rotating disk}\label{subsec:disk}

\begin{figure}
\begin{minipage}{0.49\textwidth}
\begin{center}
\includegraphics[width=1.\textwidth]{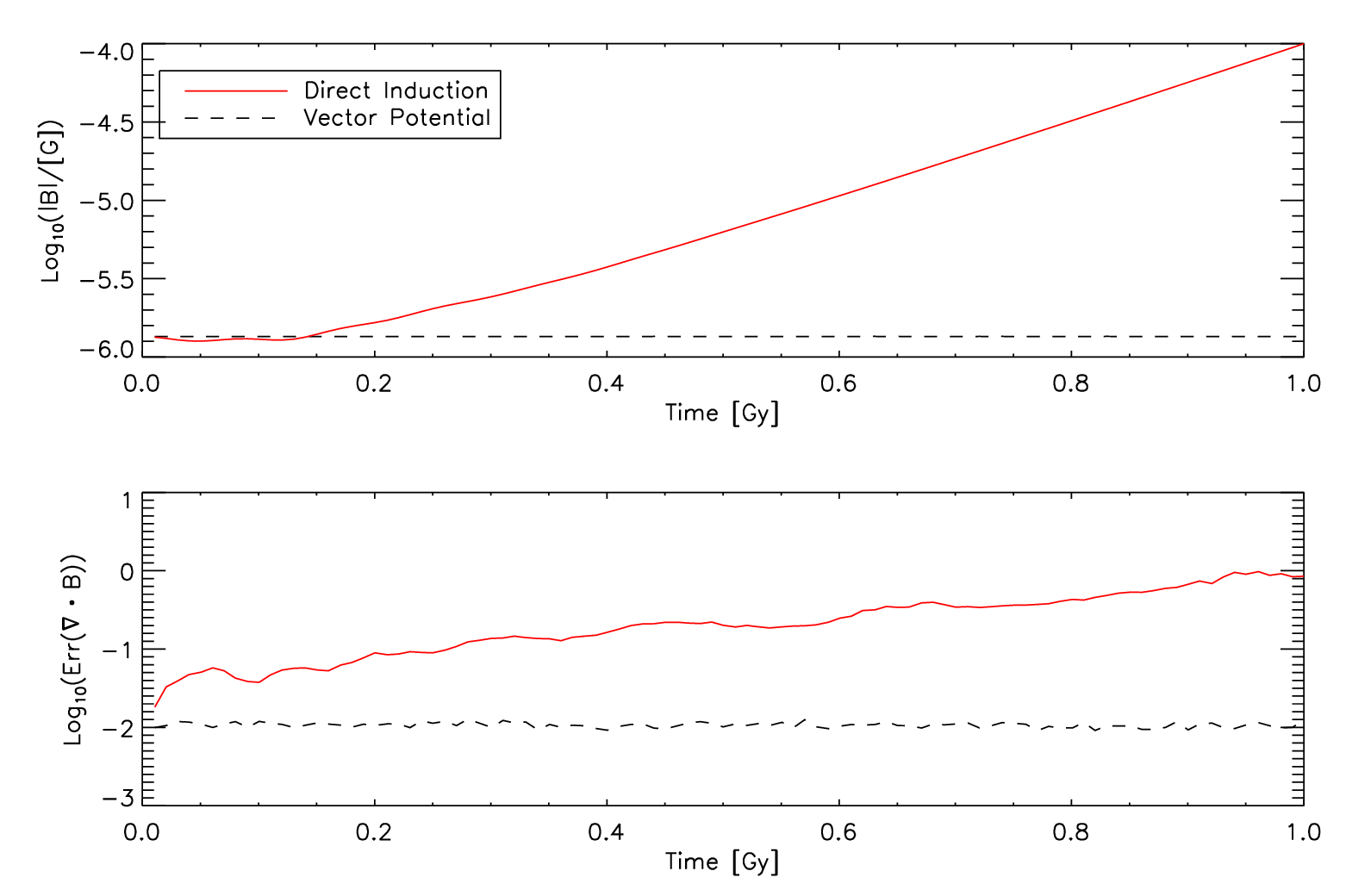}
\end{center}
\caption[Disk Evolution]
{The upper panel shows the time evolution of the average magnetic field strength of the differentially rotating disk. 
We show the evolution of the vector potential implementation (dashed-black line)
and of the direct integration (solid-red line). 
The lower panel shows the mean evolution of the divergence error that accumulates over time for the direct integration 
and stay constant in the case of the vector potential. 
}
\label{fig:DiscEvol0}
\end{minipage}
\qquad
\begin{minipage}{0.49\textwidth}
\begin{center}
\includegraphics[width=0.8\textwidth]{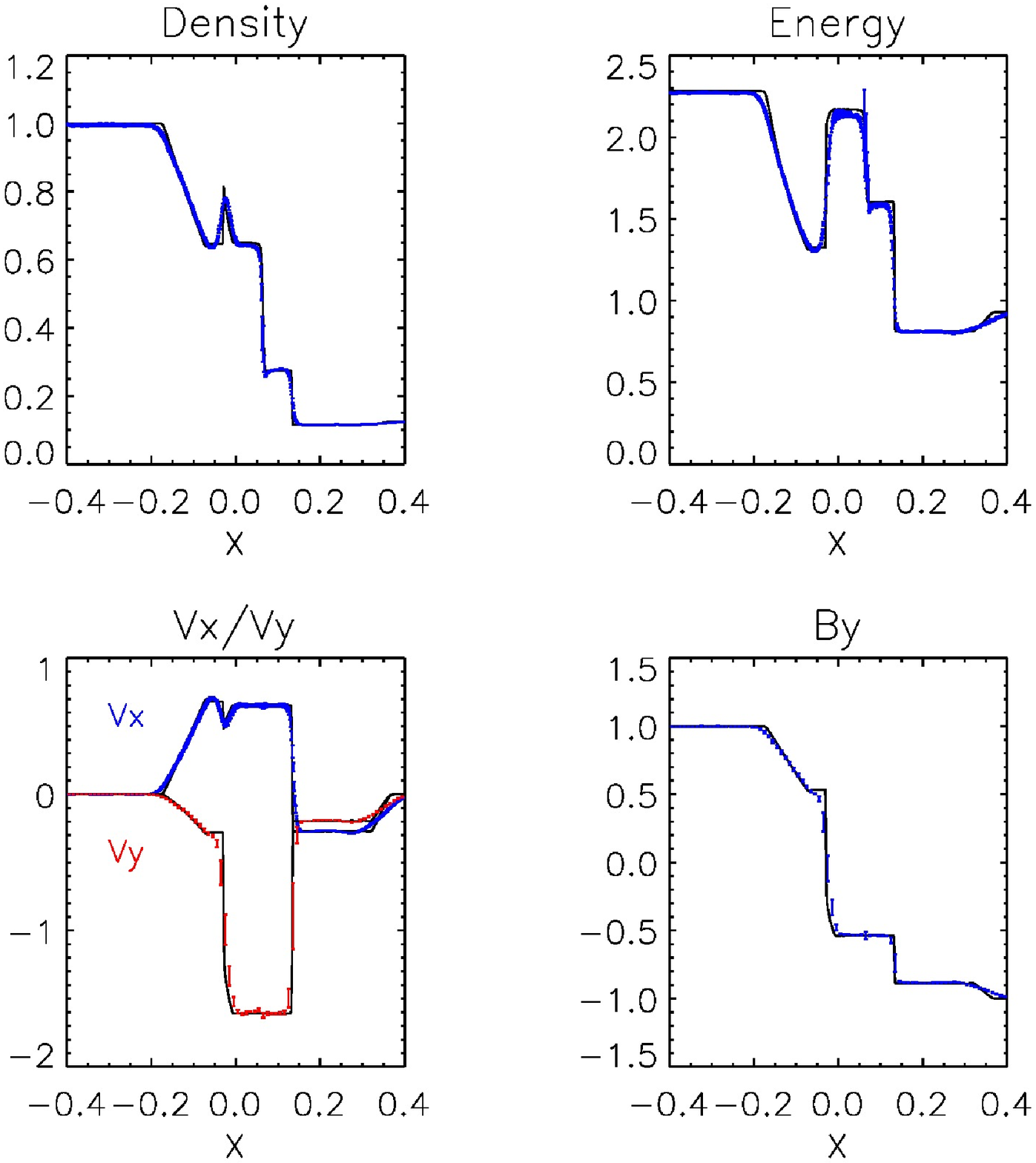}
\caption[Tube 5A]
{
In blue and red the result of different quantities of the \citet{BrioWu88} test compared with the solution from 
{\small Athena} \citep{Stone2008} in black line. Compare with \citet{PriceVector2010} and \citet{StasyszynDolag2013}, 
for other solutions. 
}
\label{fig:Tube001}
\end{center}
\end{minipage}
\end{figure}

A gravitational bound disk with differential rotation is a natural test bed for starting kinematic galactic 
magnetohydrodynamics and dynamo action. 
Magnetic field vectors parallel to the velocity vectors are stationary solutions of the induction equation for ideal MHD.   
Therefore, we setup a rotating disk, 
following a {\it Brandt} profile in the velocity distribution $V_\phi(r) = r~\Omega_0/\sqrt{1+r^2/{r_0}^2}$,
using as parameters for the model $\Omega_0=180 ~{\rm Gyr}^{-1}$ and $r_0=2.0~{\rm kpc}$.
We will omit units and interpret length normalized to 1~kpc and time to 1~Gyr.  
Additionally, we prescribe the static gravitational potential that gives equilibrium between 
the centrifugal and gravitational forces. 
Neglecting hydrodynamic forces keeps the rotation constant in time. 
We use  220949 particles in a disk with radius $R=5$ and height $H=4$.
The particles are set up initially equidistant in radius, height and polar angle but with a random phase distribution. 
Periodic boundary conditions are applied. 
We start with only a toroidal magnetic field $B_\phi$, that is the result of a vector potential 
$A_z = 0$ if $r<1$, $A_z = -0.1 \cdot (r-1.0)$ if $1<r<2$ and $A_z =-0.1$ if $r>2$. 
Because magnetic field vectors parallel to the velocity vectors are stationary solutions of the induction equation for ideal MHD, 
the field should not change during time evolution.

This simple example was numerical unstable for the direct implementation of the magnetic field but stable 
for the vector potential. 
The instability is due to the creation of a small radial magnetic field component from the {\it toroidal} field by
the discretization error. Together with the rotational shear (amplifying again
the toroidal field) this leads to an exponential growth of the field with $\divB \neq 0$.
The problematic transfer of errors between components happens
because the term $(\bfB\cdot\nabla)\bfv$ is used in \EQ{induc0} instead of the advection term $(\bfv\cdot\nabla)\bfB$.
The cleaning does not help here, because it does not remove the $B_r$ instead it adds a $B_z$, which indeed reduces 
the $\divB$, but the total field still growth spuriously by the shear and could misleadingly interpreted as dynamo action.

In \FIGp{DiscEvol0} we plot the evolution of the mean magnetic field, in dashed black lines the vector potential implementation 
and in solid red the direct induction formulation. 
In the lower panel of \FIGp{DiscEvol0} we show the evolution of the mean divergence error $Err(\divB) = h~\divB / |\bfB|$.
In the case of the direct induction we observe an exponential growth, while during the integration with the vector potential the small 
error related to the accuracy of the curl and div operator remains constant. 

\subsection{Roberts Flow}\label{subsec:Roberts}

\begin{figure}[t]
\centering
\includegraphics[width=0.99\textwidth]{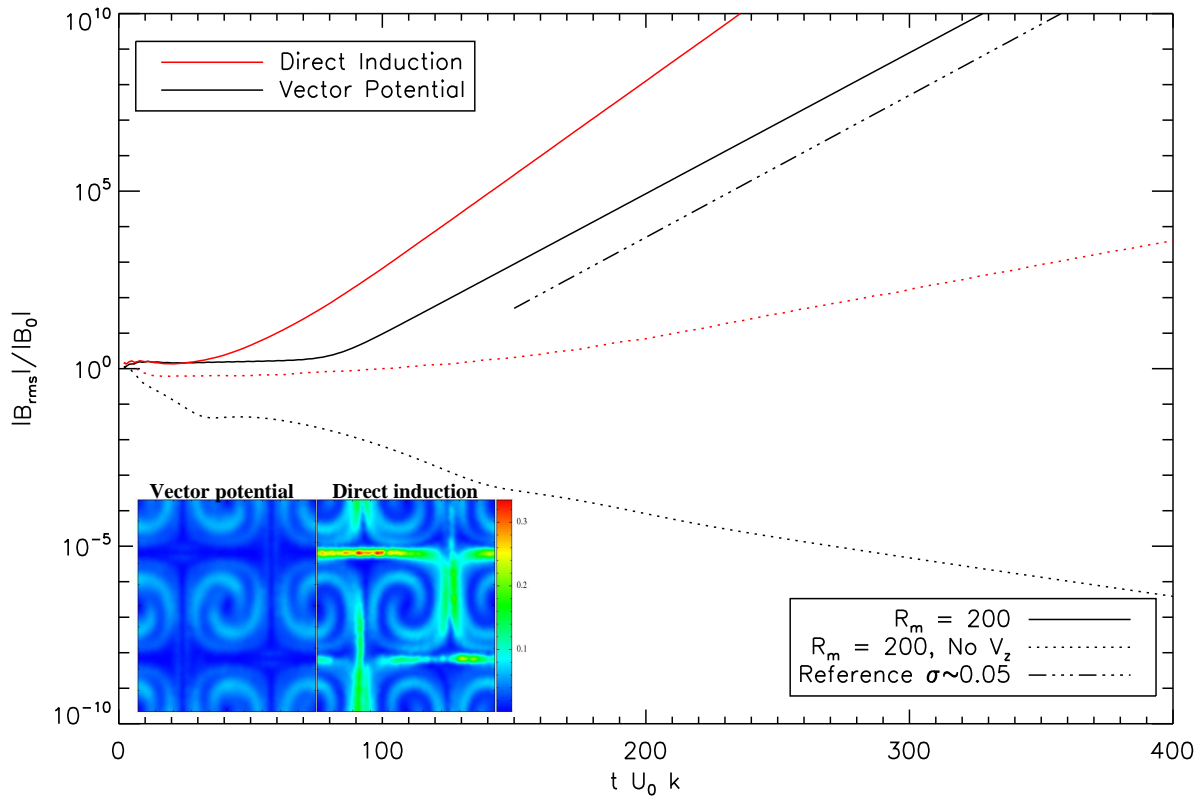}
\caption[RobertsFlow]{
Evolution of the mean magnetic energy density for a Roberts flow $R_m=200$ in different cases.
We recover the expected growth rate $\sigma \sim 0.05$ for the vector potential, as well the decay for the planar flow ($V_z = 0$).
The growth rate is larger in the direct induction implementation and a wrong growth of the magnetic field appears in the planar case.
Additionally, we show 2 cuts of the magnetic field strength for the $V_z =0$ at early times, where we already can appreciate differences in the evolution.
In the dynamo case we do not see major morphological differences. 
}
\label{fig:Roberts}
\end{figure}
 
We use the Roberts flow to demonstrate the viability of the code to handle basic dynamo case.
The periodic flow is set-up with a flow in the form
\bea
\bfV = U_0 ~ \left[ - \sin(k y) \cos(k x) , \sin(k x) \cos(k y) ,(\sqrt{2})^{-1} \cos(k y) \cos(k x) \right]
\eea{RobVel}
being $U_0$ the characteristic velocity of the flow, $k = 2~\pi/L$ and $L$ is the size of the periodic box.
Following \citet{2010MNRAS.401..347B} we set up the problem with magnetic Reynolds number $R_m=v~k^{-1}~\eta^{-1}=200$, which gives a growth rate $\sigma \sim 0.05$. 
In particular, we use a $U_0=16\pi$ and $\eta=0.04$, in code units.
To be able to do this test in SPH we set up a cubic lattice 
with $50^3$ isothermal gas particles of sufficient temperature by prescribing the sound speed $c_s=55$, which gives a maximum mach number $M=0.9$.
We let the system evolve with the velocity prescribed by the definition \EQ{RobVel} and 
seed some magnetic field (defined from the vector potentials 
\footnote{In particular we use a vector potential defined as $A_x=0$, $A_y= (2~k)^{-1}~sin(2~k~y)~cos(2~k~z)$ and $A_z=(2~k)^{-1}~sin(2~k~z)~cos(2~k~y)$} in both cases)
after a small relaxation time, when the particle distribution stabilises.
As a complement, we run tests setting the vertical velocity $V_z = 0$.
In \FIG{Roberts} we show the resulting evolution of the mean magnetic energy density in the box for the different cases, 
inlaid two cuts of the magnetic field at early stages for the $V_z = 0$ case.  
The ones with a three dimensional velocity field develops a dynamo independent of the implementation 
and with similar morphologies (therefore not shown). 
However, the growth rate in the direct induction case is larger than expected. 
In the planar case, we only have decaying solutions for the magnetic field. 
The direct induction case shows a wrong growth of the magnetic field. 
The reason seems to be the wrong advection of the field in curved flows similar to the example of section \ref{subsec:disk}.
This shows that only the scheme with the vector potential converges to the correct solution.

\subsection{Brio-Wu shock tube}\label{subsec:tube}

Basically, we use the same glass like tube setup as described in \citet{DolagStasyszyn2009},
but enlarging it 10 times in the z-direction, by replicating the initial conditions in that direction.
We set up the initial vector potential as $A_z = |x-L_X/2.|$ and $A_y = -0.75 ~ z$, 
where $L_x$ is the longitude of the tube domain in the $x$ dimension and we use a total of 350000 particles.
The particular definition of $A_y$ is the reason why we extend the shock dimension in $z$ to have good spatial derivatives.
In the code we take care that the periodic boundary conditions are well fulfilled (particularly in $z$). 
Note that we do not use an external field as \citet{PriceVector2010} for the $B_x$ component.
In \FIGp{Tube001} we show in blue the result of the test for the different quantities compared with the solution from 
{\small Athena} \citep{Stone2008} in black line. 
Overall, we found a good agreement between the solutions and previous SPMHD implementations. 
However there is a noticeable smoothing of the magnetic field, which is expected from the scheme itself.

\subsection{Orzang-Tang Vortex}\label{subsec:vort}

This test was introduced by \citet{Orzang79} and has the complexity of many kinds of interacting waves.
We use the same initial conditions as described in \citet{StasyszynDolag2013}. 
We configure the vector potential as 
$A_z = (\pi \sqrt{4~\pi})^{-1}~ [0.5~ \cos(2~ \pi~ y) + 0.25~ \cos(4~ \pi~ x)]$. 
\citet{PriceVector2010} uses a similar procedure and reports tensile instabilities when running in 3 dimensions.
We also observe instabilities at $t\approx0.4$ even with smoothing of the magnetic field, but 
we are able to avoid them by keeping track of the gauge.
At later times ($t>0.6$) instabilities rise again which is consistent with \citet{2011PhPl...18a2903C},
and on this time we are able to handle them with a little bit of diffusion 
(see section \ref{subsec:diff} and using a value in code units of $\eta = 0.0001$
\footnote{The dissipation value, is chosen to have diffusive times larger enough, compared with other timescales in the problem.}).
Therefore, we are able to run the test until late times (i.e. $t>6.0$) without any problems. Note that this test is usually presented only up to  $t\approx0.5$. 
The effect of the small dissipation does not trigger any instabilities (see also the discussion in \citet{StasyszynDolag2013}).

\begin{figure}
\begin{minipage}{0.49\textwidth}
\centering
\includegraphics[width=0.8\textwidth]{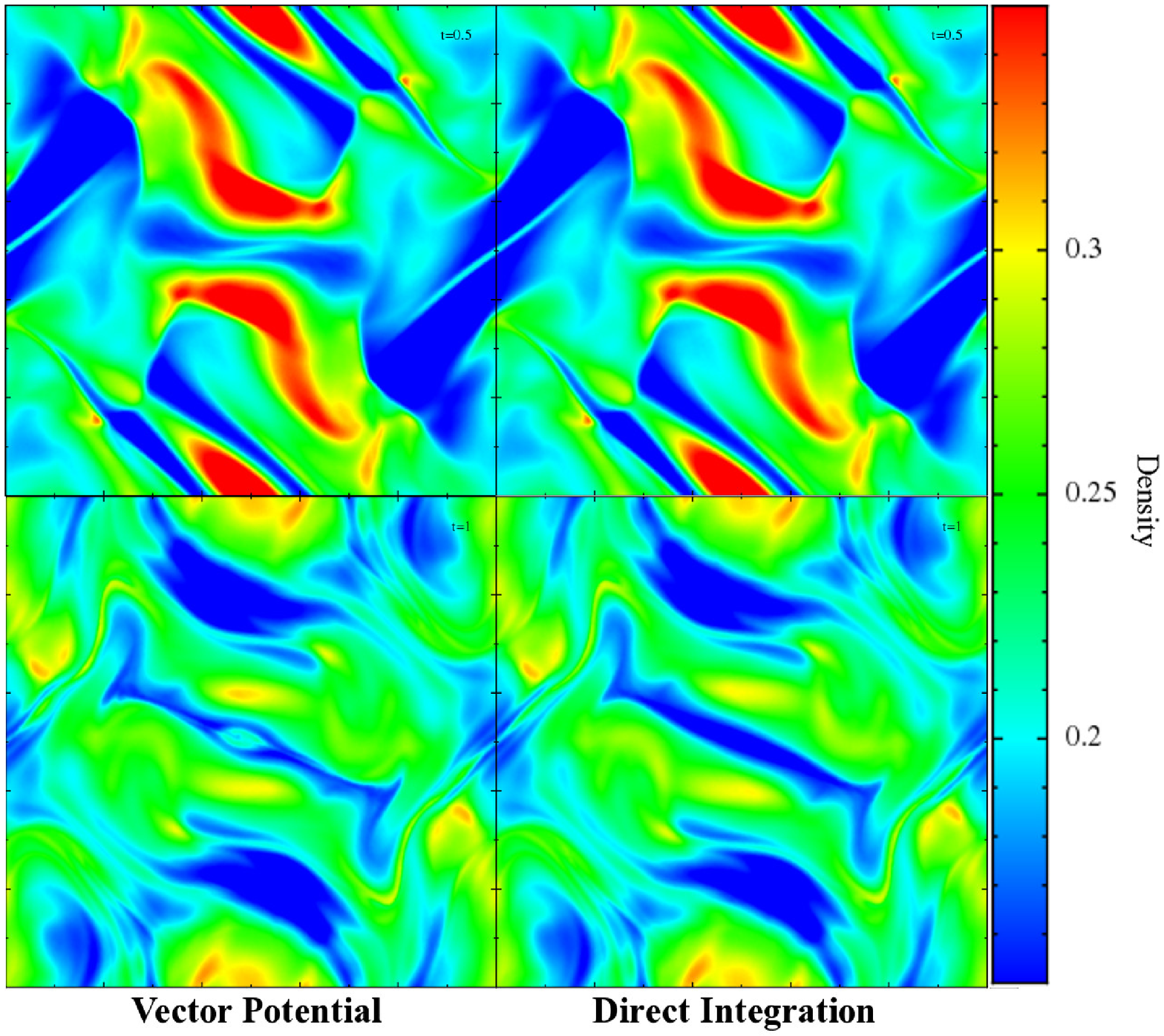}
\caption[Vortex]
{
Shown are the density for the Orzang-Tang vortex at time $t=0.5$ (upper panels) and $t=1.0$ (lower panels).
The solutions are completely comparable with previous methods. 
}
\label{fig:Vort001}
\end{minipage}
\qquad
\begin{minipage}{0.49\textwidth}
\centering
\includegraphics[width=0.95\textwidth]{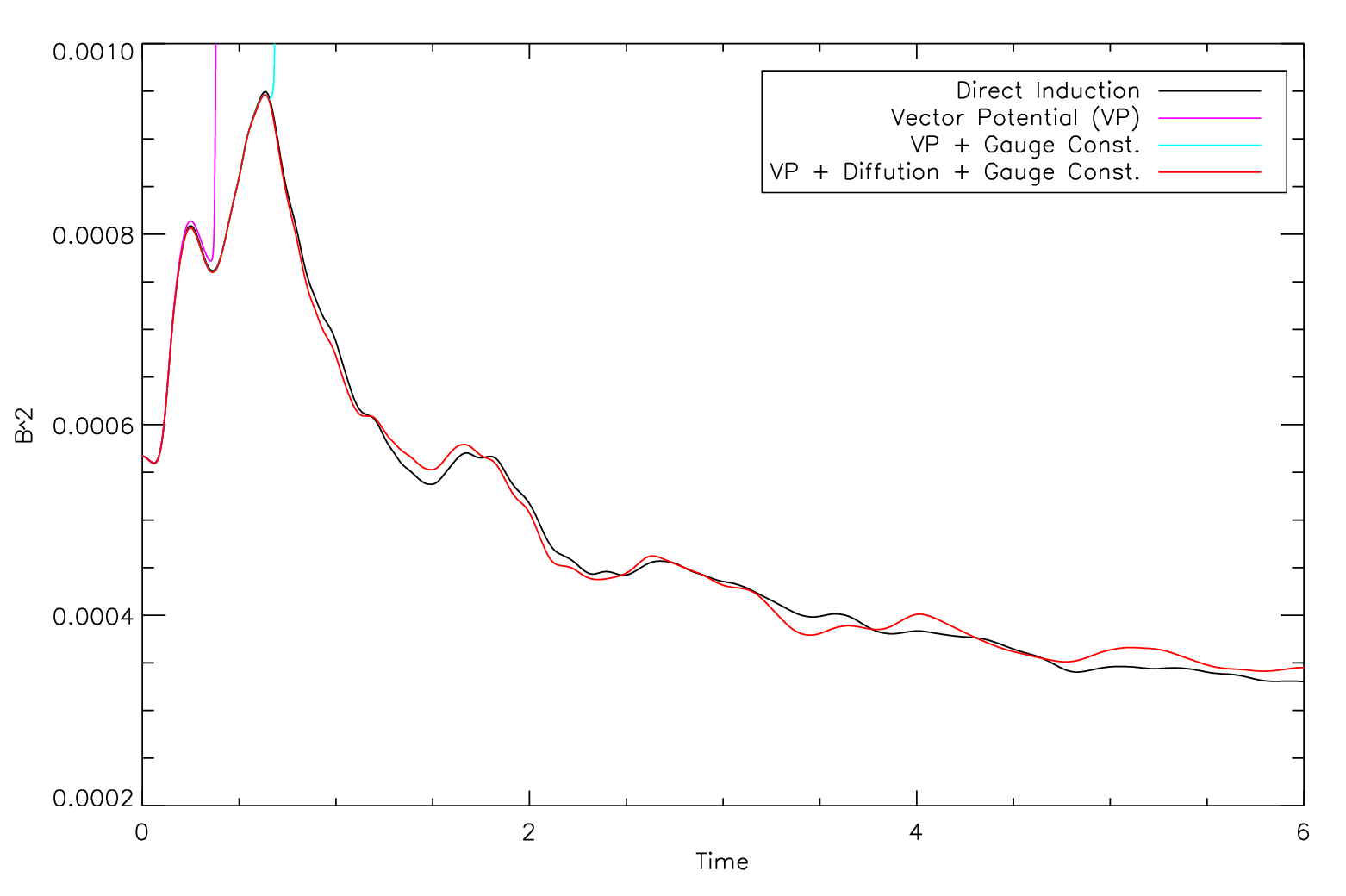}
\caption[VortEvol]
{
Evolution of the mean magnetic energy density for the Orzang-Tang vortex 
for the SPH induction implementation and the vector potential.
The introduction of the numerical corrections avoid instabilities, 
not showing noticeable differences affecting the evolution.
}
\label{fig:Vort002}
\end{minipage}
\end{figure}

In \FIGp{Vort001}, we show the density distribution for two different times, 
$t=0.5$ (which is the one usually used to compare between codes) and $t=1.0$.
The solutions found are completely comparable between them and with other methods. 
In \FIGp{Vort002}, we show the evolution of the magnetic energy density for these different implementations. 
They match quite well and the symmetry of the problem is overall maintained. 

\section{Implications}\label{sec:disc}

The sole fact that the usual implementation fail in the simple test of section \ref{subsec:disk} is worrisome.
The code integrates all the variables in cartesian coordinates, which is not optimal for our cylinder symmetric test 
with a dominant rotational velocity. But Eulerian grid methods can handle this problem easily in cartesian coordinates. 
Rotational objects are a common feature of astrophysical simulations and should be traced in a stable and consistent scheme. 
We cannot rely on the fact that the stochastic approach of the method can cancel out all the errors, but this should be demonstrated. 
In particular, this test points to the danger to confuse a numerical instability with a physical dynamo instability. 
Therefore it should be taken into account in the interpretation of the results of the simulations. We exemplify the problem with 
an astrophysical model of a galaxy.

\subsection{Example: A Galaxy}\label{subsec:gala}

To corroborate possible implications of the methods, we compare both methods in an astrophysical scenario with a fully 3 dimensional flow. 
We set up a small galaxy, with $3.9\times10^4$ SPH particles in disk and IGM, and $3.4\times10^5$ DM particles. 

\begin{figure}[t]
\centering
\includegraphics[width=0.85\textwidth]{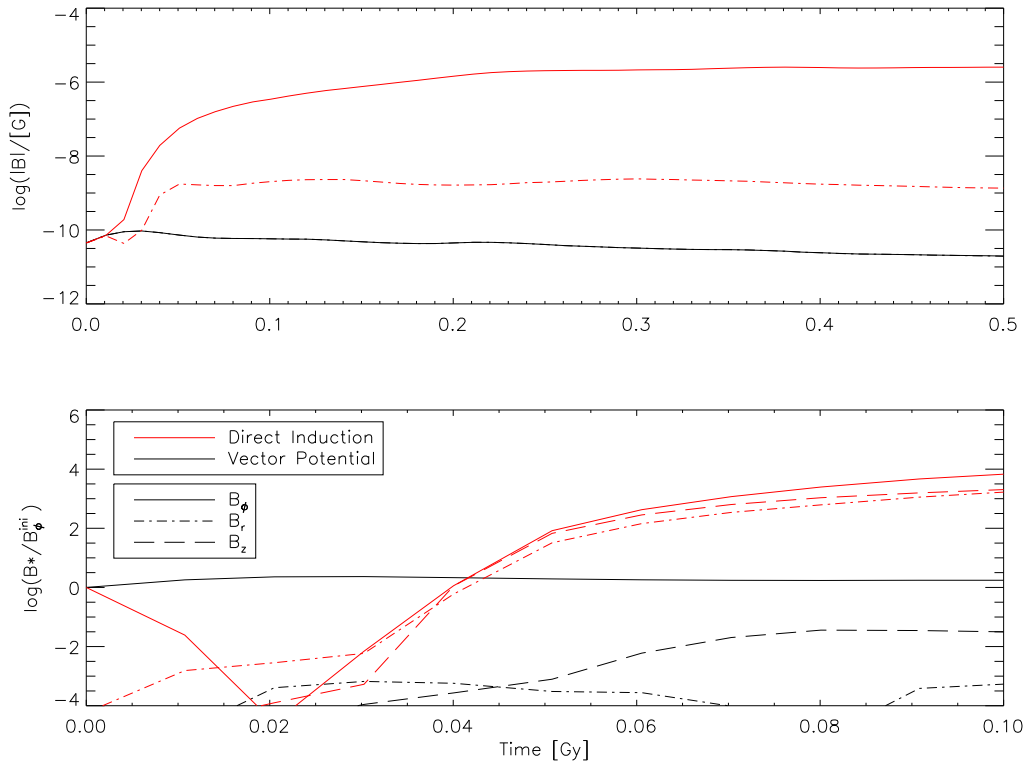}
\caption[GalaEvol]
{
In the upper panel we show the evolution of the mean magnetic field strength for the different methods (red - direct integration, black - vector potential).
Solid lines represent the result of the kinematic case (without Lorentz force) and dot dashed lines of the full MHD implementation.
In the lower panel we show the evolution of the absolute value of the different components averaged in cylindrical coordinates 
and normalized to the initial $B_\phi$ (solid line is $B_\phi$, long dashed is $B_z$ and dot dashed is $B_r$) for the kinematic case. 
}
\label{fig:GalaEvol}
\end{figure}

The SPH particles are distributed in a disk and inter-stellar medium to avoid spurious effects from the boundary \citep{Annette2012}.
The total mass of the galaxy is $2.4\times10^{11} \mathrm{M\odot}$ and the initial radius of the disk is $r_0=20~\kpc$.
We let the galaxy evolve during $0.2~\Gyr$, to allow the particles to relax and stabilize, after this time we seed a 
magnetic field in the disk by a ring of $B_\phi = 10^{-12} \mathrm{G}$ calculated using only a vector potential in the $A_z$ 
component in the same way as described in section \ref{subsec:disk}. 
We also try a constant $B_x$ as initial magnetic configuration with similar results.
We evolve the galaxy with both schemes using a small constant dissipation $\eta~ = ~6\times10^{24} \mathrm{cm^2/sec}$.
Again the direct integration of the magnetic field shows a strong growth of the magnetic energy in contrast to the potential method, 
where the field slowly decays (cf. \FIGp{GalaEvol}).
We consider also the dynamical cases, taking into account the Lorentz force.

In the upper panel of \FIGp{GalaEvol}, we show the evolution of the mean magnetic energy for the different methods.
In red we plot the direct induction and the vector potential in black.
Also, solid lines are the result without taking into account the Lorentz force and in dot dashed the full MHD implementation.
In the beginning of the simulation both methods seem to agree, afterwards there is a growth to $\muG$ levels in the induction case.
In the lower panel we show the log absolute value of the average for the different components 
in cylindrical coordinates normalized to the initial $B_\phi$ for an initial subset of the time evolution, until $0.1~\Gyr$. 
This is shown only for the kinematic case, therefore the gas dynamics is the same, but the magnetic fields evolution differ. 
The direct integration leads to a strong numerical growth of the poloidal magnetic field components $B_r$ and $B_z$ 
in a similar way as found in the test problem of section \ref{subsec:disk}. 

\begin{figure}[t] 
\centering
\includegraphics[width=0.9\textwidth]{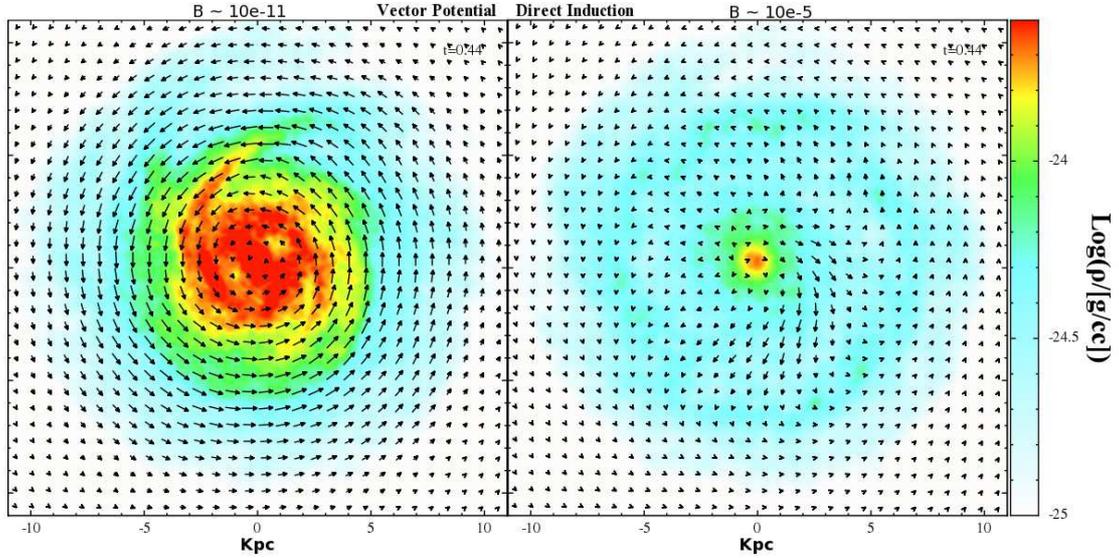}
\caption[Gala0]
{
Density cuts through the disk of the galaxy in the dynamic simulations with vector potential (left) 
and direct integration of the magnetic field (right).
Superimposed are the magnetic field vectors normalized to $10^{-11}G$ and $10^{-5}G$, respectively.
}
\label{fig:Gala0}
\end{figure}

For the dynamical simulations, we show density cuts in \FIGp{Gala0} with their respective magnetic field vectors, 
which illustrate the difference caused probably by the unphysical growth of the magnetic field for the direct induction algorithm. 
The magnetic field remains weak with no dynamical influence on the gas for the solution with the vector potential in contrast
to the direct integration of the induction equation  with locally 6 orders of magnitude larger field strength.
The density is in the same log scale, showing that in the induction case the magnetic pressure 
prevents the accretion of gas, meanwhile in the vector potential case, we reach higher densities and stronger spiral arms, 
and there is almost no difference in the evolution with the kinematic case. 

\citet{Kotarba2009}  found a similar behavior for a comparison between Euler potentials and direct induction implementations 
applied to an isolated galaxy. 
In that case, a stronger growth of the magnetic field in the center of the galaxy was found for the direct induction case 
when compared to the Euler potential, which is consistent with our results. 
\citet{Kotarba2009} justify the difference in terms of $\divB$ errors, and discard the full reliability of the Euler implementation 
because of the fact that the winding up of the field is not correctly traced by the Euler potentials while they are 
just advected with the velocity field. 
This is also demonstrated by \citet{StasyszynDolag2013}, in which the \citet{Orzang79} vortex is evolved several winding up times, and the system evolved through Euler potentials turn out to be unstable at late times.
In our case, the vector potentials are fully evolved, which is found to be reliable even in a galaxy scenario.

\section{Conclusions} \label{sec:conc}

This work was triggered by the failing tests of the induction equation, 
which are naturally solved with the vector potential formalism. This example demonstrates that a 
divergence error cleaning method does not guarantee to recover the {\it consistent} soleonidal solution. 
We were able to build a vector potential implementation of the SPMHD equations,
that successfully passes test cases and an astrophysical scenario.
However, more testing and studies are needed, but escapes the scope of the current exploratory work.

In summary, we evolve the vector potential, from which we calculate the magnetic field.
We smooth the resulting magnetic field to regularize it and avoid tensile instabilities 
in the force calculation. 
We found that the constraint of a correct evolution of the gauge is also important, 
and we implement a solution similar to the divergence cleaning from \citet{StasyszynDolag2013} 
but applied to the vector potential field to be consistent with a pseudo-Coulomb gauge.
There are still several possibilities for the gauges, that could be studied in the future.
The smoothing of the field and the special care of the gauge in the potential evolution 
seem to be the key points for which previous implementations failed.
Additionally, we implement an explicit diffusion to the equation of the vector potential.  
This helps to stabilize tests, allowing for example to run the \citet{Orzang79} vortex up to at least $6$ winding times.
The kinematic dynamo of the Roberts flow gave the expected growth rate for a 3D velocity field and a finite magnetic diffusivity, 
as well in the planar case. The same test fail in both cases for the direct induction with growing solutions.

The comparison of both methods for the astrophysical application of the galaxy evolution shows a similar
behavior as the simple rotating disk example. A probably unphysical growth of the initial toroidal magnetic field 
appears only for the direct integration, while the vector potential method leads only to a radial redistribution of the toroidal
field.

Solving directly the induction equation in SPH has been applied in the past.
We understand that the standard implementation of the induction does not advect correctly the field in some
cases. Such effect has an unclear net effect in the stochastic motions of astrophysical simulations, 
being possible to be washed out, or not.

The use of the hybrid approach in order to couple the dynamics and evolution of the magnetic field from the 
vector potential opens space to improve the numerical implementation. 
The same has to be said for the additional dissipative switches, energy conservation and a deeper study of the method in several 
environments, which already have been performed in the case of the direct induction evolution in SPMHD.
However, the implementation presented here is already robust enough for further applications investigating dynamo processes in astrophysics. 


\section*{Acknowledgements}

We  thank Klaus Dolag for the useful discussions and the referees 
for their comments that helped to improve this work.
For the revision of the text we are grateful to Creasey P. and Lagos C. and Paz D, 
and comments of the Roberts flow form Candelaresi S. and Del Sordo F.
Part of this work was supported by the Deutsche Forschungsgemeinschaft (DFG), project number FOR1254.
Figure 3 and 6 have been done using {\small SPLASH} \citep{2007PASA...24..159P}.

\section*{References}
\bibliographystyle{elsarticle-harv}
\bibliography{master}

\appendix

\end{document}